\documentclass[10pt]{article}
\usepackage{amssymb,amsbsy}
\usepackage{amsmath}
\usepackage{amsthm}
\usepackage{graphics,graphicx}
\usepackage[T1]{fontenc}
\usepackage{color}
\newcommand{\beq}{\begin{equation}}
\newcommand{\eeq}{\end{equation}}

\newcommand{\beqar}{\begin{eqnarray}}
\newcommand{\eeqar}{\end{eqnarray}}
\newcommand{\bit}{\begin{itemize}}
\newcommand{\eit}{\end{itemize}}
\newcommand{\benum}{\begin{enumerate}}
\newcommand{\eenum}{\end{enumerate}}
\newcommand{\barr}{\begin{array}}
\newcommand{\earr}{\end{array}}
\def\ds{\displaystyle}
\newcommand\eq[1]{(\ref{#1})}
\newcommand{\jump}[2]{[\mbox{\hspace{-#1em}}[#2]\mbox{\hspace{-#1em}}]}
\newcommand{\modIII}{\text{III}}

\def\XXint#1#2#3{{\setbox0=\hbox{$#1{#2#3}{\int}$}
   \vcenter{\hbox{$#2#3$}}\kern-.5\wd0}}

\def\b0{\mbox{\boldmath $0$}}

\def\bc{\mbox{\boldmath $c$}}

\def\be{\mbox{\boldmath $e$}}

\def\bn{\mbox{\boldmath $n$}}

\def\bx{\mbox{\boldmath $x$}}

\def\bI{\mbox{\boldmath $I$}}

\def\bP{\mbox{\boldmath $P$}}

\def\bY{\mbox{\boldmath $Y$}}

\newcommand{\bxi}{\mbox{\boldmath $\xi$}}

\newcommand{\bPhi}{\mbox{\boldmath $\Phi$}}

\def\f0{\ensuremath{\mathbb{O}}}

\newcommand{\bmM}{\mbox{\boldmath $\mathcal{M}$}}

\def\Re{\mathop{\mathrm{Re}}}

\newcommand{\Reals}{\ensuremath{\mathbb{R}}}

\def\EFM{{\it Eng.\ Fract.\ Mech.}\ }

\def\IJF{{\it Int.\ J.\ Fracture}\ }

\def\IJSS{{\it Int.\ J.\ Solids Struct.}\ }

\def\JAM{{\it ASME J.\ Appl.\ Mech.}\ }
\def\JAP{{\it J.\ Appl.\ Phys.}\ }

\def\JMPS{{\it J.\ Mech.\ Phys.\ Solids}\ }

\def\MOM{{\it Mech.\ Materials}\ }

\title{Perturbation analysis of Mode III interfacial cracks advancing in a dilute heterogeneous material}

\author{A. Piccolroaz$^{(1)}$, G. Mishuris$^{(2)}$,
A.B. Movchan$^{(3)}$, N. Movchan$^{(3)}$
\\
\\
$^{(1)}$
{\it Dipartimento di Ingegneria Meccanica e
Strutturale, Universit\`a di Trento,}
\\ {\it Via Mesiano 77, I-38050 Trento, Italia}
\\
$^{(2)}$
{\it Institute of Mathematical and Physical Sciences, Aberystwyth
University, }
\\ {\it Ceredigion SY23 3BZ, Wales U.K.,}
\\
$^{(3)}$
{\it Department of Mathematical Sciences, University of Liverpool, }
\\ {\it Liverpool L69 3BX, U.K.}
}

\begin{document}

\maketitle

\begin{abstract}
\noindent
The paper addresses the problem of a Mode III interfacial crack advancing quasi-statically in a heterogeneous composite material, that is a two-phase material 
containing elastic inclusions, both soft and stiff, and defects, such as microcracks, rigid line inclusions and voids. It is assumed that the bonding between 
dissimilar elastic materials is weak so that the interface is a preferential path for the crack. The perturbation analysis is made possible by means of the 
fundamental solutions (symmetric and skew-symmetric weight functions) derived in Piccolroaz et al. (2009). We derive the dipole matrices of the defects in 
question and use the corresponding dipole fields to evaluate effective tractions along the crack faces and interface to describe the interaction between the main 
interfacial crack and the defects. For a stable propagation of the crack, the perturbation of the stress intensity factor induced by the defects is then 
balanced by the elongation of the crack along the interface, thus giving an explicit asymptotic formula for the calculation of the crack advance. 
The method is general and applicable to interfacial cracks with general distributed loading on the crack faces, taking into account possible asymmetry in 
the boundary conditions.

The analytical results are used to analyse the shielding and amplification effects of various types of defects in different configurations. Numerical computations 
based on the explicit analytical formulae allows for the analysis of crack propagation and arrest.
\end{abstract}

\newpage

\tableofcontents

\newpage

\section{Introduction}
\label{sec1}

Analytic solutions for a crack propagating in a {\it homogeneous} elastic solid containing a finite number of small defects (elastic and rigid inclusions, 
microcracks, voids) have been derived in Bigoni et al. (1998), Valentini et al. (1999), Movchan et al. (2002) on the basis of the dipole matrix and weight 
function approach. The dipole field describes the leading-order perturbation produced by a small defect placed in a smooth stress field and gives rise to 
``effective'' tractions applied along the crack faces. Furthermore, the weight functions allow for the derivation of the corresponding perturbation of the 
stress intensity factor as weighted integral of the ``effective'' loading. The method is general and applicable to both 2D and 3D cases and to defects of 
different type and shape, provided that the corresponding dipole matrix is appropriately constructed and the weight functions for the corresponding unperturbed 
cracked body are available.

Problems of a macrocrack interacting with microcracks have been analysed by Romalis and Tamuzh (1984) under the influence of mechanical loading and by Tamusz 
et al. (1993) under the influence of heat flux, using the analytic functions and singular integrals approach (Muskhelishvili, 2008). The possible closure of 
crack surfaces and the consequent appearance of a contact zone have been considered in Tamuzs et al. (1994) and Tamuzs et al. (1996). Asymptotic models of a 
semi-infinite crack interacting with microcracks have been developed by Gong and Horii (1989) and Meguid et al. (1991) using complex potentials and the 
superposition principle. A review of publications on macro--microcrack interaction problem is given in Tamuzs and Petrova (2002).

Recently, Piccolroaz et al.\ (2009) derived the symmetric and skew-symmetric weight functions for a semi-infinite two-dimensional interfacial crack, thus 
disclosing the possibility of applying the dipole matrix approach to the propagation of cracks along the interface in heterogeneous materials with small defects.
The weight functions constructed in Piccolroaz et al.\ (2009) are of the generalised type and thus applicable to any type of boundary conditions along the 
crack faces. The singular perturbation associated with a small crack advance was also obtained there. It is worth noting that the 
availability of the skew-symmetric function for the problem under consideration is essential since the ``effective'' loading produced by the defects on the 
crack faces is not symmetrical, in general. In the present paper, we analyse the scalar case of antiplane shear loading. The full vector problem will be 
addressed elsewhere.

The paper is organised as follows. The formulation of the problem is outlined in Section \ref{sec2}, which includes also preliminary results on the unperturbed 
problem. Section \ref{sec3} is devoted to the perturbation analysis, in particular the derivation of the dipole fields for several types of defects and the 
analysis of singular perturbation associated with the crack advance. Section \ref{sec4} provides a number of numerical results based on the explicit analytical 
formulae. In particular, shielding and amplification effects of different defect configurations on the crack-tip field are presented. The possibility to design 
a neutral configuration for any given force system distributed along the crack faces is discussed. Finally, the crack propagation and arrest produced by the 
defects under consideration is analysed. In the appendix, we derive the dipole matrix for an elliptic elastic inclusion placed in a homogeneous antiplane field.

\section{Problem formulation and preliminary results}
\label{sec2}
\begin{figure}[!htcb]
\begin{center}
\includegraphics[width=11cm]{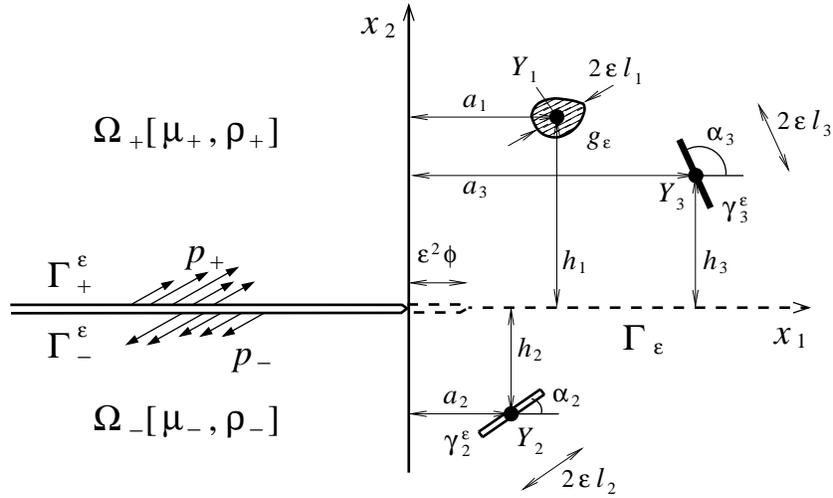}
\caption{\footnotesize Geometry of the problem: interface crack in a bimaterial plane with defects: $g_\varepsilon$ denotes a small elastic inclusion, 
$\gamma_2^\varepsilon$ a microcrack, and $\gamma_3^\varepsilon$ a rigid line inclusion; $\bY_1,\bY_2,\bY_3$ are the ``centres'' of the defects.}
\label{fig01}
\end{center}
\end{figure}

We consider a two-dimensional composite structure consisting of a bimaterial matrix (two dissimilar elastic half-planes $\Omega_\pm$) containing a dilute 
distribution of inclusions, microcracks and rigid line inclusions, see Fig.~\ref{fig01}. The two materials constituting the matrix are assumed to be linear 
elastic and isotropic, with shear moduli denoted by $\mu_+$ and $\mu_-$, respectively. All interfaces between different phases are assumed to be perfect, 
that is, the displacements and tractions remain continuous across the interface.

We introduce the following notations. Let $g_\varepsilon \subset \Omega_+$ be a small elastic inclusion of diameter $2\varepsilon l_1$ centred at the point 
$\bY_1 = (a_1,h_1)$. The shear modulus of the inclusion is denoted by $\mu_i$, and it can be greater or smaller than the  shear modulus $\mu_+$ of the 
surrounding material, so that both stiff and soft inclusions are considered. The notation $\gamma_2^\varepsilon \subset \Omega_-$ is used for a microcrack of 
the length $2 \varepsilon l_2$, centred at the point $Y_2 = (a_2,h_2)$ and making an angle $\alpha_2 > 0$ with the positive direction of the $x_1$-axis. 
By $\gamma_3^\varepsilon \subset \Omega_+$ we denote a movable rigid line inclusion of the length $2 \varepsilon l_3$, centred at the point $Y_3 = (a_3,h_3)$ and 
making an angle $\alpha_3 > 0$ with the positive direction of the $x_1$-axis. Although these notations refer specifically to Fig.\ \ref{fig01}, the 
formulation can be easily extended to problems with different number and type of defects, as those considered in Sec.\ \ref{sec4}.

We assume that a semi-infinite interface crack $M_\varepsilon$ advances quasi-statically along the interface $\Gamma_\varepsilon$ connecting the half-planes, and 
we denote the uniform advance of the crack by $\varepsilon^2 \phi$. Here and in the sequel, $\varepsilon > 0$ is a small dimensionless parameter. The reason for 
the order $\varepsilon^2$ in the crack advance will be clear in Sec.\ \ref{sec3}, where we perform the asymptotic analysis.
%%%%%%%%%%%%%%%%%%%%%%%%%%%%%%%%%%%%%%%%%%%%%%%%%%%%%%%%%%%%%%%%%%%%%%
%%%%%%%%%%%%%%%%%%%%%%%%%%%%%%%%%%%%%%%%%%%%%%%%%%%%%%%%%%%%%%%%%%%%%%

We assume that the composite is dilute, that is, the small defects are distant from each other so that the interaction between them can be neglected. 
Consequently, we can model the three cases of an elastic inclusion, of a microcrack and of a rigid line inclusion separately. It is possible then to 
extend the results to a finite number of defects by superposition using the linearity of the problem, provided that the distance between defects remains 
finite.

An external loading $p_\pm$ is applied to the crack faces $\Gamma_\pm^\varepsilon$ and it is assumed to be self-balanced such that the principal force vector 
is zero, that is,
\beq
\int_{\Gamma_+^\varepsilon} p_+ d\bx - \int_{\Gamma_-^\varepsilon} p_- d\bx = 0.
\eeq
We assume that the loading $p_\pm$ on the crack faces vanishes in a neighbourhood of the crack tip.

The problem is then formulated in terms of the Laplace equation
\beq
\label{laplace}
\Delta u_\pm(x_1,x_2) = 0, \quad \Delta u_i(x_1,x_2) = 0,
\eeq
where $u = \{u_+,u_-,u_i\}$ denotes the displacement component along $x_3$-axis in the respective domain 
$\Omega_+ \setminus \{g_\varepsilon \cup \gamma_3^\varepsilon\}$, $\Omega_- \setminus \gamma_2^\varepsilon$ and $g_\varepsilon$.

We prescribe the following boundary conditions on the crack faces
\beq
\label{boundary_crack}
\mu_\pm \frac{\partial u_\pm}{\partial x_2} = p_\pm \quad
\text{on $\Gamma_\pm^\varepsilon$},
\eeq
and the interface conditions
\beq
\label{interface}
u_+ = u_-, \quad
\mu_+ \frac{\partial u_+}{\partial x_2} = \mu_- \frac{\partial u_-}{\partial x_2} \quad
\text{on $\Gamma^\varepsilon$}.
\eeq
The transmission conditions for the elastic inclusion $g_\varepsilon$ are formulated similarly to (\ref{interface}), that is, 
\beq 
\label{inclusion} 
u_+ = u_i, \quad \mu_+
\frac{\partial u_+}{\partial n} = \mu_i \frac{\partial
u_i}{\partial n} \quad \text{on $\partial g_\varepsilon$}.
\eeq
We assume that the microcrack faces $\gamma_2^\pm$ are traction free, so that 
\beq 
\label{microcrack} 
\frac{\partial u_-}{\partial n} = 0 \quad
\text{on} \quad \gamma_2^\pm. 
\eeq
Finally, on the boundary of the {\em movable} rigid line inclusion $\gamma_3^\varepsilon$ the Dirichlet boundary condition is prescribed, that is, 
\beq 
\label{rigid} 
u_+ = u_* \quad \text{on} \quad \gamma_3^\varepsilon,
\eeq 
where $u_*$ is an unknown constant which will be defined later from the balance condition 
\beq
\label{balance_rigid} 
\int_{\gamma_3^\varepsilon} \frac{\partial u_+}{\partial n} ds = 0.
\eeq

The unperturbed problem ($\varepsilon = 0$) corresponds to a semi-infinite interfacial crack in a bimaterial plane. The solution to this problem (see, 
for example, Piccolroaz et al. 2010) can be expressed in the form 
\beq
\label{tildeu0}
u_\pm^{(0)}(r,\theta) = \frac{1}{2\pi i} \int_{\omega - i\infty}^{\omega +i\infty}
\tilde{u}_\pm^{(0)}(s,\theta) r^{-s} ds, \quad \pm \theta \in (0,\pi),
\eeq
where
\beq
\tilde{u}^{(0)}_\pm(s,\theta) =
-\frac{\sin(s\theta)}{\mu_\pm s \cos(\pi s)} \langle \tilde{p} \rangle(s) +
\left[ \frac{\cos(s\theta)}{(\mu_+ + \mu_-) s \sin(\pi s)} +
\frac{(\mu_+ - \mu_-) \sin(s\theta)}{2\mu_\pm (\mu_+ + \mu_-) s \cos(\pi s)} \right] 
\jump{0.15}{\tilde{p}}(s),
\eeq
and $0 < \omega < 0.5$, with the symmetric and skew-symmetric components of the loading denoted by $\jump{0.15}{p} = p_+ - p_-$ and 
$\langle p \rangle = (p_+ + p_-)/2$, respectively. 

The two-terms asymptotic expansions for the traction $\sigma^{(0)}$ ahead of the crack tip and for the crack opening $\jump{0.15}{u^{(0)}}$ read 
as follows
\beq
\sigma^{(0)}(r,0) = \frac{K^{(0)}_\modIII}{\sqrt{2\pi}} r^{-1/2} + \frac{A^{(0)}_\modIII}{\sqrt{2\pi}} r^{1/2} + O(r^{3/2}),
\eeq
\beq
\jump{0.15}{u^{(0)}}(r) =
\frac{\mu_+ + \mu_-}{\mu_+\mu_-} \left(\frac{2K^{(0)}_\modIII}{\sqrt{2\pi}} r^{1/2} - \frac{2A^{(0)}_\modIII}{3\sqrt{2\pi}} r^{3/2}\right) + O(r^{5/2}),
\eeq
where
\beq
\label{K_30}
K^{(0)}_\modIII = -\sqrt{\frac{2}{\pi}} \int_0^\infty \left\{ \langle p \rangle(-r) + \frac{\eta}{2} \jump{0.15}{p}(-r) \right\} r^{-1/2} dr,
\eeq
\beq
\label{A_30}
A^{(0)}_\modIII = \sqrt{\frac{2}{\pi}} \int_0^\infty \left\{ \langle p \rangle(-r) + \frac{\eta}{2} \jump{0.15}{p}(-r) \right\} r^{-3/2} dr,
\eeq
with $\eta = (\mu_- - \mu_+)/(\mu_+ + \mu_-)$ being the contrast parameter.

For the perturbation analysis we will also need the so-called symmetric and skew-symmetric weight functions, $\jump{0.15}{U}$ and $\langle U \rangle$, 
respectively, and the corresponding traction $\langle \Sigma \rangle$ in a bimaterial plane with a semi-infinite crack (see Piccolroaz et al., 2009),
\beq
\label{wfIII}
\jump{0.15}{U}(x_1) =
\left\{
\barr{ll}
\ds \frac{1-i}{\sqrt{2\pi}} x_1^{-1/2}, & \ds x_1 > 0, \\[3mm]
0, & x_1 < 0,
\earr
\right. \quad
\langle U \rangle(x_1) = \eta/2 \jump{0.15}{U}(x_1),
\eeq
\beq
\label{wfIII2}
\langle \Sigma \rangle(x_1) =
\left\{
\barr{ll}
0, & \ds x_1 > 0, \\[3mm]
\ds \frac{(1-i) \mu_+\mu_-}{2\sqrt{2\pi}(\mu_+ + \mu_-)} (-x_1)^{-3/2}, & x_1 < 0.
\earr
\right.
\eeq

For an arbitrary solution $u$ of the governing equation (\ref{laplace}) and the corresponding traction $\sigma$ along the $x_1$-axis, the reciprocity 
identity has the form (Piccolroaz et al., 2009)
\beq
\label{recid}
\int_{-\infty}^\infty \Big\{ \jump{0.15}{U}(x_1'-x_1) \langle \sigma \rangle(x_1) + 
\langle U \rangle(x_1'-x_1) \jump{0.15}{\sigma}(x_1)
- \langle \Sigma \rangle(x_1'-x_1) \jump{0.15}{u}(x_1) \Big\} dx_1 = 0,
\eeq
where the following notations have been used:
$$
\jump{0.15}{f} = f_+ - f_-, \quad \langle f \rangle = \frac{1}{2}(f_+ + f_-).
$$

Let us introduce the notations
$$ 
f^{(+)}(x_1) = f(x_1) H(x_1), \quad f^{(-)}(x_1) = f(x_1) H(-x_1), 
$$
where $H$ denotes the Heaviside function, so that
\beq
f(x_1) = f^{(+)}(x_1) + f^{(-)}(x_1).
\eeq

The reciprocity identity \eq{recid} can be written as
\beq
\label{recid2}
\barr{l}
\ds
\int_{-\infty}^{\infty} \left\{
\jump{0.15}{U}(x_1'-x_1) \langle \sigma \rangle^{(+)}(x_1)
- \langle \Sigma \rangle(x_1'-x_1) \jump{0.15}{u}^{(-)}(x_1)
\right\} dx_1 =
\\[3mm]
\ds
- \int_{-\infty}^{\infty} \left\{
\jump{0.15}{U}(x_1'-x_1) \langle \sigma \rangle^{(-)}(x_1)
+ \langle U \rangle(x_1'-x_1) \jump{0.15}{\sigma}^{(-)}(x_1)
\right\} dx_1
\\[3mm]
\ds
- \int_{-\infty}^{\infty} \langle U \rangle(x_1'-x_1)
\jump{0.15}{\sigma}^{(+)}(x_1) dx_1
+ \int_{-\infty}^{\infty} \langle \Sigma \rangle(x_1'-x_1)
\jump{0.15}{u}^{(+)}(x_1) dx_1.
\earr
\eeq

Note that $\langle \sigma \rangle^{(-)}$, $\jump{0.15}{\sigma}^{(-)}$ are the average and the jump of the prescribed loading on the crack faces, 
whereas $\jump{0.15}{\sigma}^{(+)}$, $\jump{0.15}{u}^{(+)}$ are the prescribed discontinuities of traction and displacement across the interface.
The identity (\ref{recid2}) will be used extensively later in the perturbation analysis.

Additionally, we shall need the components of the displacement gradient $\nabla_{\bx} u_\pm^{(0)}$ computed at an arbitrary point 
$\bY = (d\cos\varphi, d\sin\varphi)$. These are obtained from (\ref{tildeu0}) and given by
\beq
\barr{r}
\ds \left.\frac{\partial u_\pm^{(0)}}{\partial x_1}\right|_{\bY} = \frac{1}{\pi d} \int_{-\infty}^0 \frac{dx_1}{2\cos\varphi - x_1/d - d/x_1} 
\left\{\frac{\jump{0.15}{p}(x_1)}{\mu_+ + \mu_-} \left[\vphantom{\sqrt{\frac{-x_1}{d}}} \sin^2\varphi - \frac{1}{2}\cos\varphi\left(\frac{d}{x_1} - 
\frac{x_1}{d}\right)\right] + \right. \\[5mm]
\ds \left. + \frac{2\langle p \rangle(x_1) + \eta\jump{0.15}{p}(x_1)}{2\mu_\pm} \left(\sqrt{\frac{-x_1}{d}}\sin\frac{\varphi}{2} + 
\sqrt{\frac{d}{-x_1}}\sin\frac{3\varphi}{2}\right)\right\},
\earr
\eeq
\beq
\label{nabla}
\barr{r}
\ds \left.\frac{\partial u_\pm^{(0)}}{\partial x_2}\right|_{\bY} = -\frac{1}{\pi d} \int_{-\infty}^0 \frac{dx_1}{2\cos\varphi - x_1/d - d/x_1} 
\left\{\frac{\jump{0.15}{p}(x_1)\sin\varphi}{\mu_+ + \mu_-} \left[\vphantom{\sqrt{\frac{-x_1}{d}}} \cos\varphi + \frac{1}{2}\left(\frac{d}{x_1} - 
\frac{x_1}{d}\right)\right] + \right. \\[5mm]
\ds \left. + \frac{2\langle p \rangle(x_1) + \eta\jump{0.15}{p}(x_1)}{2\mu_\pm} \left(\sqrt{\frac{-x_1}{d}}\cos\frac{\varphi}{2} + 
\sqrt{\frac{d}{-x_1}}\cos\frac{3\varphi}{2}\right)\right\}.
\earr
\eeq

\section{Perturbation analysis}
\label{sec3} 

We shall construct an asymptotic solution of the problem using the method of Movchan and Movchan (1995), that is the asymptotics of the solution will 
be taken in the form
\beq 
\label{exp}
u_\pm(\bx,\varepsilon) = u_\pm^{(0)}(\bx)  + \varepsilon \sum_{j=1}^{3} W_j(\bxi_j) + 
\varepsilon^2 \sum_{j=1}^{3} u_\pm^{(j)}(\bx) + \varepsilon^2 v(\bx,\phi) + o(\varepsilon^2), \quad \varepsilon \to 0. 
\eeq
In (\ref{exp}), the leading term $u_\pm^{(0)}(\bx)$ corresponds to the unperturbed solution, and it is described in the previous section. The term 
$\varepsilon \sum_{j=1}^{3} W_j(\bxi_j)$ corresponds to the boundary layers concentrated near the defects and needed to satisfy the respective conditions 
(\ref{inclusion}), (\ref{microcrack}) and (\ref{rigid}); the variables $\bxi_j$ will be defined in the next section. The term 
$\varepsilon^2 \sum_{j=1}^{3} u_\pm^{(j)}(\bx)$ is introduced to fulfil the original boundary conditions (\ref{boundary_crack}) on the crack faces and the 
interface conditions (\ref{interface}) disturbed by the boundary layers; this term, in turn, will produce perturbations of the crack tip fields and 
correspondingly of the stress intensity factor. In Sec.\ \ref{sec31} we will analyse the effect of each defect separately. Finally, the term 
$\varepsilon^2 v(\bx,\phi)$ corresponds to a singular perturbation associated with a possible crack advance, $\varepsilon^2 \phi$, and will be addressed in 
Sec.\ \ref{sec32}.

\subsection{Singular perturbation and dipole fields generated by small defects}
\label{sec31}

\subsubsection{Small elastic inclusion}
\label{sec311}
We shall start with the elastic inclusion, situated in the upper half-plane. The leading term $u^{(0)} = u^{(0)}_+$ clearly does not satisfy the transmission 
conditions (\ref{inclusion}) on the boundary $\partial g_\varepsilon$. Thus, we shall correct the solution by constructing the boundary layer $W_1(\bxi_1)$, 
where the new scaled variable $\bxi_1$ is defined by
\beq
\label{Y1}
\bxi_1 = \frac{\bx - \bY_1}{\varepsilon},
\eeq
with $\bY_1$ being the ``centre'' of the inclusion $g_\varepsilon$ (see Fig. \ref{fig01}).

For $W_1(\bxi_1) = \{W_1^{in}, \ \bxi_1 \in g; \ W_1^{out}, \ \bxi_1 \in \Reals^2 \setminus \overline{g}\}$ we consider the following problem 
\beq
\label{laplace1} 
\Delta W_1^{in}(\bxi_1) = 0, \quad \bxi_1 \in g, \quad \Delta W_1^{out} = 0, \quad 
\bxi_1 \in \Reals^2 \setminus \overline{g},
\eeq 
where
$$
g = \varepsilon^{-1} g_\varepsilon \equiv
\{\bxi_1 \in \Reals^2: \varepsilon\bxi_1 + \bY_1 \in g_\varepsilon\}.
$$
The function $W_1$ remains continuous across the interface $\partial g$, that is, 
$$
W_1^{in} = W_1^{out} \quad \text{on} \quad \partial g,
$$
and satisfies on $\partial g$ the following transmission condition 
\beq
\mu_i \frac{\partial}{\partial \bn} W_1^{in}(\bxi_1) -
\mu_+ \frac{\partial}{\partial \bn} W_1^{out}(\bxi_1) =
(\mu_+ - \mu_i) \bn \cdot \nabla u^{(0)}(\bY_1) + O(\varepsilon), \quad \varepsilon \to 0,
\eeq
where $\bn = \bn_{\bxi_1}$ is an outward unit normal on $\partial g$. The formulation is completed by setting the following condition at infinity
\beq
\label{infinity}
W_1^{out} \to 0 \quad \text{as} \quad |\bxi_1| \to \infty.
\eeq
The problem above has been solved by various techniques and the solution can be found, for example, in Movchan and Movchan (1995).

Since we assume that the inclusion is at a finite distance from the interface between the half-planes, we shall only need the leading term of the 
asymptotics of the solution at infinity. This term reads as follows
\beq 
\label{boundary_layer_1}
W_1^{out}(\bxi_1) = -\frac{1}{2\pi}
\left[\left.\nabla_{\bx}\, u^{(0)}\right|_{\bY_1}\right] \cdot 
\left[\bmM_1 \frac{\bxi_1}{|\bxi_1|^2}\right] + 
O(|\bxi_1|^{-2}) \quad \text{as} \quad \bxi_1 \to
\infty, 
\eeq 
where $\bmM_1$ is a 2 $\times$ 2 matrix which depends on the characteristic size $l_1$ of the domain $g$ and the ratio $\mu_+/\mu_i$; it is called the dipole 
matrix. For example, in the case of an elliptic inclusion with the semi-axes $l_a$ and $l_b$ making an angle $\alpha_1$ with the positive direction of the $x_1$-axis 
and $x_2$-axis, respectively, the matrix $\bmM_1$ takes the form
\beq
\label{dip}
\bmM_1 = -\frac{\pi}{2} l_a l_b (1 + e)(\mu_\star - 1)
\left[
\barr{ll}
\ds 
\frac{1 + \cos 2\alpha_1}{e + \mu_\star} + \frac{1 - \cos 2\alpha_1}{1 + e\mu_\star} &
\ds 
-\frac{(1 - e)(\mu_\star - 1) \sin 2\alpha_1}{(e + \mu_\star)(1 + e\mu_\star)} \\[3mm]
\ds 
-\frac{(1 - e)(\mu_\star - 1) \sin 2\alpha_1}{(e + \mu_\star)(1 + e\mu_\star)} & 
\ds 
\frac{1 - \cos 2\alpha_1}{e + \mu_\star} + \frac{1 + \cos 2\alpha_1}{1 + e\mu_\star} 
\earr
\right],
\eeq
where $e = l_b/l_a$ and $\mu_\star = \mu_+/\mu_i$. We note that for a soft inclusion, $\mu_+ > \mu_i$, the dipole matrix is negative definite, whereas 
for a stiff inclusion, $\mu_+ < \mu_i$, the dipole matrix is positive definite.

The term $\varepsilon W_1(\bxi_1)$ in a neighbourhood of the $x_1$-axis written in the $\bx$ coordinates takes the form
\beq 
\label{BL_1_crack_faces_1} 
\varepsilon W_1(\bxi_1) = \varepsilon^2
w_1(\bx) + o(\varepsilon^2), \quad \varepsilon \to 0,
\eeq
where
\beq 
\label{BL_1_crack_faces_2} 
w_1(\bx) = -\frac{1}{2\pi} \left[\left.\nabla_{\bx} u^{(0)}\right|_{\bY_1}\right] \cdot 
\left[\bmM_1 \frac{\bx - \bY_1}{|\bx - \bY_1|^2}\right]. 
\eeq

As a result, one can compute the average $\varepsilon^2 \langle \sigma_1 \rangle$ and the jump $\varepsilon^2 \jump{0.15}{\sigma_1}$ of the ``effective'' traction 
across the line $x_2 = 0$ induced by the elastic inclusion $g_\varepsilon$. 
Since $\partial u^{(1)}/\partial x_2 = -\partial w_1/\partial x_2$ must hold on the line $x_2 = 0$ (to satisfy the original boundary conditions 
(\ref{boundary_crack}) and interface conditions (\ref{interface})), this gives
\beq 
\label{traction_discrepancy_1} 
\langle \sigma_1 \rangle = 
-\frac{1}{2}(\mu_+ + \mu_-) \frac{\partial w_1}{\partial x_2}, \quad 
\jump{0.15}{\sigma_1} = 
-(\mu_+ - \mu_-) \frac{\partial w_1}{\partial x_2},
\eeq
where
\beq
\label{dw1dx2}
\frac{\partial w_1}{\partial x_2} = 
-\frac{1}{2\pi} \left[\left.\nabla_{\bx} u^{(0)}\right|_{\bY_1}\right] \cdot 
\bmM_1 \frac{\be_2}{|\bx - \bY_1|^2} + 
\frac{1}{\pi} \left[\left.\nabla_{\bx} u^{(0)}\right|_{\bY_1}\right] \cdot 
\bmM_1 \frac{(\bx - \bY_1)(x_2 - Y_{12})}{|\bx - \bY_1|^4}
\eeq

In the limit $\mu_i \to \infty$, we obtain the dipole matrix for a rigid movable inclusion. In the case of an elliptic rigid inclusion, we have
\beq
\label{rigid-ellipse}
\bmM = \frac{\pi}{2} l_a l_b (1/e + 1)
\left[
\barr{ll}
1 + \cos 2\alpha + e(1 - \cos 2\alpha) & (1 - e)\sin 2\alpha \\[3mm]
(1 - e)\sin 2\alpha & 1 - \cos 2\alpha + e(1 + \cos 2\alpha)
\earr
\right].
\eeq

\subsubsection{Microcrack or small void}
\label{sec312}

Now we apply the same procedure to a microcrack. Since the leading term $u^{(0)}$ does not satisfy traction free boundary conditions (\ref{microcrack}) on 
the microcrack faces $\gamma_2^\pm$, we construct a boundary layer $W_2(\bxi_2)$, where the new scaled variable $\bxi_2$ is defined similarly to (\ref{Y1}) 
with $\bY_1$ replaced by $\bY_2$, the middle point of the microcrack (see Fig. \ref{fig01}).

The function $W_2$ satisfies the Laplace equation in a homogeneous plane containing a crack of finite length $2l_2$ and the following traction conditions 
on the crack faces
\beq
\frac{\partial}{\partial \bn} W_2(\bxi_2) =
- \bn \cdot \nabla u^{(0)}(\bY_2) + O(\varepsilon), \quad \varepsilon \to 0,
\eeq
where $\bn = \bn_{\bxi_2}$ is an outward unit normal on the crack faces of the finite crack 
$\gamma_2 = \varepsilon^{-1} \gamma_2^\varepsilon \equiv \{\bxi_2 \in \Reals^2: \varepsilon\bxi_2 + \bY_2 \in \gamma_2^\varepsilon\}$.
Again, we are looking for the boundary layer solution which decays at infinity. Such a problem has been solved by various techniques and 
the solution can be found, for example, in Muskhelishvili (2008). The asymptotic formulae (\ref{boundary_layer_1}), (\ref{BL_1_crack_faces_1}) and 
(\ref{BL_1_crack_faces_2}) remain the same with the dipole matrix $\bmM_1$ replaced by
\beq
\label{crack}
\bmM_2 = -\pi l_2^2
\left[
\barr{ll}
\sin^2\alpha_2 & -\sin\alpha_2\cos\alpha_2 \\[3mm]
-\sin\alpha_2\cos\alpha_2 & \cos^2\alpha_2
\earr
\right] = 
-\frac{\pi l_2^2}{2}
\left[
\barr{ll}
1 - \cos 2\alpha_2 & -\sin 2\alpha_2 \\[3mm]
-\sin 2\alpha_2 & 1 + \cos 2\alpha_2
\earr
\right],
\eeq
giving the average $\varepsilon^2 \langle \sigma_2 \rangle$ and the jump $\varepsilon^2 \jump{0.15}{\sigma_2}$ of traction across the interface between the two 
half-planes in the form of (\ref{traction_discrepancy_1}) with $w_1$ replaced by $w_2$.

For the case of a small void, one can construct the corresponding boundary layer following Movchan et al.\ (2002). It has been shown that for an 
arbitrary finite void, the dipole matrix coincide with the one corresponding to an ``equivalent'' elliptic void. Denoting by $l_a$ and $l_b$ the semi-axes 
of the ellipse, and by $\alpha_2$ the angle formed by $l_a$-semi-axis and the $x_1$-axis, the dipole matrix can be written as follows
\beq
\label{ellipse}
\bmM = -\frac{\pi}{2}(l_a + l_b)^2 
\left\{ 
\bI + \frac{1-e}{1+e}
\left[
\barr{ll}
-\cos 2\alpha & -\sin 2\alpha \\[3mm]
-\sin 2\alpha & \cos 2\alpha
\earr
\right]
\right\},
\eeq
where $e = l_b/l_a$ is the eccentricity of the ellipse. One can see that in the limit case as $e \to 0$, the formula (\ref{crack}) is recovered from 
(\ref{ellipse}).

\subsubsection{Rigid line inclusion}
\label{sec313}

In the case of a small rigid line inclusion, the boundary layer solution $W_3(\bxi_3)$ satisfies the Laplace equation (\ref{laplace1}) in the entire plane 
of the scaled coordinates system $\bxi_3\in \Reals^2$ introduced as in (\ref{Y1}), where the point $\bY_1$ is replaced by $\bY_3$ and stands for the middle 
point of the rigid inclusion (see Fig.\ \ref{fig01}).

On the boundary of the rigid inclusion $\gamma_3 = \varepsilon^{-1} \gamma_3^\varepsilon \equiv \{\bxi_3 \in \Reals^2: \varepsilon\bxi_3 + \bY_3 
\in \gamma_3^\varepsilon\}$ the function $W_3$ satisfies the following condition (compare with (\ref{rigid})):
$$
\varepsilon W_3(\bxi_3) = u_* - u^{(0)}(\bY_3)- \varepsilon \left.\nabla_{\bx}
u^{(0)}\right|_{\bY_3} \cdot \bxi_3 + O(\varepsilon^2), \quad \varepsilon \to 0.
$$
Substituting (\ref{exp}) into (\ref{rigid}) gives
\beq
u_* = u_+^{(0)}(\bY_3) + \varepsilon \left.\nabla_{\bx} u_+^{(0)}\right|_{\bY_3} \cdot \bxi_3 + 
\varepsilon W_3(\bxi_3) + O(\varepsilon^2), \quad \varepsilon \to 0.
\eeq
To satisfy this condition we need to set $u_* = u_+^{(0)}(\bY_3)$ and 
\beq 
\label{w3}
W_3(\bxi_3) = -\left.\nabla_{\bx} u^{(0)}\right|_{\bY_3} \cdot \bxi_3, \quad \bxi_3 \in \gamma_3. 
\eeq
Finally, we are looking for a solution which decays at infinity similarly to (\ref{infinity}).

Such a problem can been solved by various techniques. We shall present the construction of the dipole field using complex analysis in 
\ref{app02}, following Muskhelishvili (2008). The corresponding dipole matrix takes the form
\beq
\label{rigid-line}
\bmM_3 =  
\frac{\pi l_3^2}{2}
\left[
\barr{ll}
1 + \cos 2\alpha_3 & \sin 2\alpha_3 \\[3mm]
\sin 2\alpha_3 & 1- \cos 2\alpha_3
\earr
\right].
\eeq
One can see that in the limit case as $e \to 0$, the formula (\ref{rigid-line}) is recovered from (\ref{rigid-ellipse}).

In the original coordinate system $\bx$, far away from the rigid inclusion, the asymptotics of the boundary layer solution $W_3$ takes the same form as 
(\ref{BL_1_crack_faces_1}) and (\ref{BL_1_crack_faces_2}), where $w_1$ and $\bmM_1$ should be replaced by $w_3$ and $\bmM_3$, respectively.

The average and jump of traction across the interface crack can be found by formulae analogous to (\ref{traction_discrepancy_1}).

\subsubsection{Line defect with imperfect bonding}
\label{sec314}

The dipole field and corresponding dipole matrix for a line defect with imperfect bonding can be derived, as a limiting case, from the solution for a 
perfectly bonded elastic elliptic inclusion, see Sec. \ref{sec311}. We distinguish between stiff and soft line defect.

The transmission conditions for a soft line defect (Antipov et al., 2001; Mishuris, 2001; Hashin, 2001) are given by
\beq
\label{trsoft}
\jump{0.15}{\sigma}(s) = 0, \quad \jump{0.15}{u}(s) = \kappa \sigma(s),
\eeq
where $s$ is an abscissa along the line inclusion, $\jump{0.15}{u}$ and $\sigma$ are the displacement jump across the line inclusion and the respective 
(continuous) traction, and $\kappa$ is the compliance of the bonding. Correspondingly, the dipole matrix is derived from (\ref{dip}) by choosing 
$\mu_* l_b = \kappa$ and then taking the limit $l_b \to 0$. We obtain
\beq
\label{soft}
\bmM = -\frac{\pi}{2} \frac{l^2 \kappa}{l + \kappa}
\left[
\barr{cc}
1 - \cos 2\alpha & -\sin 2\alpha \\
-\sin 2\alpha & 1 + \cos 2 \alpha
\earr
\right].
\eeq 
One can check that the solution for a microcrack (\ref{crack}) is recovered by taking the limit $\kappa \to \infty$ in (\ref{soft}).

For a stiff line defect, the transmission conditions (Benveniste and Miloh, 2001; Mishuris, 2003) are written as
\beq
\label{trstiff}
\jump{0.15}{u}(s) = 0, \quad \kappa\jump{0.15}{\sigma}(s) + \left.\frac{\partial^2 u}{\partial s^2}\right|_{\gamma} = 0.
\eeq
In this case, the dipole matrix is derived from (\ref{dip}) by choosing $\mu_* = \kappa l_b$ and then taking the limit $l_b \to 0$. We obtain
\beq
\label{stiff}
\bmM = \frac{\pi}{2} \frac{l^2}{1 + \kappa l}
\left[
\barr{cc}
1 + \cos 2\alpha & \sin 2\alpha \\
\sin 2\alpha & 1 - \cos 2 \alpha
\earr
\right].
\eeq 
One can check that the solution for a rigid line inclusion (\ref{rigid-line}) is recovered by taking the limit $\kappa \to 0$ in (\ref{stiff}).

\subsection{Singular perturbation for crack advance}
\label{sec32}

In this section, we consider a singular perturbation of the physical fields generated by an advance of the crack tip by a small quantity 
$\varepsilon^2 \phi$. We denote unperturbed quantities by subscript $0$ and perturbed quantities by subscript $\star$. We also write the perturbed fields with 
reference to a coordinate system centred at the new crack tip position.

Writing the Betti identity (\ref{recid2}) for the unperturbed ($\sigma_0$ and $\jump{0.15}{u_0}$) and the perturbed ($\sigma_\star$ and $\jump{0.15}{u_\star}$) 
quantites and then subtracting one from the other, we obtain, after the Fourier transform, (see for details Piccolroaz et al., 2009) 
\beq
\label{fou}
\jump{0.15}{\overline{U}}^+ 
(\overline{\sigma}_0^+ - e^{i\beta \varepsilon^2\phi}\overline{\sigma}_\star^+) - 
\overline{\Sigma}^-
(\jump{0.15}{\overline{u}_0}^- - 
e^{i\beta \varepsilon^2\phi}\jump{0.15}{\overline{u}_\star}^-) = 0.
\eeq
In (\ref{fou}) the bar denotes Fourier transform with respect to the variable $x_1$, and $\beta$ the corresponding variable in the transformed space. 
The superscripts $+$ and $-$ characterise functions analytic in the upper and lower half planes respectively.

Following Willis and Movchan (1995), we expand the exponential term as $\exp(i\beta \varepsilon^2\phi) = 1 + i\varepsilon^2\phi\beta + O(\varepsilon^4)$ and also 
substitute into (\ref{fou}) the two-terms asymptotic of traction $\overline{\sigma}^+_0$ and crack opening $\jump{0.15}{\overline{u}_0}^-$, as 
$\beta_\pm \to \infty$:
\beq
\label{expan}
\overline{\sigma}^+_0 = \frac{(1+i)K_\modIII^{(0)}}{2} \beta_+^{-1/2} - 
\frac{(1-i)A_\modIII^{(0)}}{4} \beta_+^{-3/2} + O(\beta_+^{-5/2}),
\eeq
\beq
\label{expan2}
\jump{0.15}{\overline{u}_0}^- = 
- \frac{(1+i)(\mu_+ + \mu_-)K_\modIII^{(0)}}{2\mu_+\mu_-} \beta_-^{-3/2} + 
\frac{(1-i)(\mu_+ + \mu_-)A_\modIII^{(0)}}{4\mu_+\mu_-} \beta_-^{-5/2} + O(\beta_-^{-7/2}).
\eeq
Note that the perturbed fields, $\overline{\sigma}^+_\star$ and $\jump{0.15}{\overline{u}_\star}^-$, have the same asymptotic expansion as in (\ref{expan}) 
and (\ref{expan2}), subject to replacing $K_\modIII^{(0)}$ with $K_\modIII^\star$ and $A_\modIII^{(0)}$ with $A_\modIII^\star$. Then, collecting like powers of 
$\beta_\pm$, we get
\beq
\label{willy}
\left\{ \frac{1+i}{2}(K_\modIII^{(0)} - K_\modIII^\star) + 
\frac{1-i}{4} i \varepsilon^2\phi A_\modIII^\star \right\}
(\beta_+^{-1} - \beta_-^{-1}) + O(\beta^{-2}) = 0.
\eeq
From (\ref{willy}) we can conclude that, in the limit as $\varepsilon \to 0$, the first-order perturbation of the stress intensity factor due to a uniform 
advance of the crack tip is given by
\beq
\label{kphi}
K_\modIII^* - K_\modIII^{(0)} := \varepsilon^2 \Delta K_\modIII^\phi = \frac{\varepsilon^2\phi}{2} A_\modIII^{(0)},
\eeq
where $A_\modIII^{(0)}$ is the coefficient in the second-order term asymptotics of the unperturbed fields and it is given in terms of the loading applied to the 
crack faces as follows
\beq
\label{a30}
A_\modIII^{(0)} = \sqrt{\frac{2}{\pi}} \int_{-\infty}^{0}
\left(\langle p \rangle(x_1) + \frac{\eta}{2} \jump{0.15}{p}(x_1)\right) (-x_1)^{-3/2} dx_1.
\eeq

Note that, although a competing double expansion, as $\varepsilon \to 0$ and $\beta_\pm \to \infty$, has been used in this procedure, the correctness of the 
final result has been proved in Piccolroaz et al. (2009) by means of the Wiener--Hopf technique.

\subsection{Analysis of a stable quasi-static propagation of an interfacial crack}
\label{sec33}

The stress intensity factor is expanded as follows
\beq
\label{expan3}
K_\modIII = K_\modIII^{(0)} + \varepsilon^2 \left(\Delta K_\modIII^\phi + 
\sum_{j=1}^{3} \Delta K_\modIII^{(j)}\right) + o(\varepsilon^2), \quad \varepsilon \to 0,
\eeq
where $K_\modIII^{(0)}$ is the stress intensity factor for the unperturbed crack, $\Delta K_\modIII^\phi$ is the singular perturbation due to the uniform 
advance of the crack tip, $\sum_{j=1}^{3} \Delta K_\modIII^{(j)}$ is the perturbation produced by the defects.

Assuming that the crack is advancing quasi-statically along the interface, the energy release rate remains constant and equal to the critical value, 
$G = G_c$, so that
\beq
\label{err}
\Delta G = G - G^{(0)} = 0.
\eeq
The energy release rate can be represented in terms of the stress intensity factor by
\beq
G = \frac{1}{4}\left(\frac{1}{\mu_+} + \frac{1}{\mu_-}\right) K_\modIII^2,
\eeq
so that (\ref{err}) becomes
\beq
K_\modIII = K_\modIII^{(0)}.
\eeq
Upon replacing in the above equation the expansion (\ref{expan3}), we obtain
\beq
2\varepsilon^2 K_\modIII^{(0)} \left(\Delta K_\modIII^\phi + \sum_{j=1}^{3} \Delta K_\modIII^{(j)}\right) + o(\varepsilon^2) = 0, \quad \varepsilon \to 0,
\eeq
and thus
\beq
\Delta K_\modIII^\phi + \sum_{j=1}^{3} \Delta K_\modIII^{(j)} = 0,
\eeq
where $\Delta K_\modIII^\phi$ is given by (\ref{kphi}), whereas $\Delta K_\modIII^{(j)}$ is given by
\beq
\label{delk}
\Delta K_\modIII^{(j)} = -\sqrt{\frac{2}{\pi}} \int_{-\infty}^{0}
\left(\langle \sigma^{(j)} \rangle + \frac{\eta}{2} \jump{0.15}{\sigma^{(j)}}\right) 
(-x_1)^{-1/2} dx_1,
\eeq
with $\langle \sigma^{(j)} \rangle$ and $\jump{0.15}{\sigma^{(j)}}$ being the average and the jump of the traction induced by the defects across the crack 
faces, given by (\ref{traction_discrepancy_1})--(\ref{dw1dx2}). The integral in (\ref{delk}) can be evaluated explicitly and we obtain the formula
\beq
\label{delk2}
\Delta K_\modIII^{(j)} = - \sqrt{\frac{2}{\pi}} \frac{\mu_+\mu_-}{\mu_+ + \mu_-} \left.\nabla_{\bx} u^{(0)}\right|_{\bY_j} \cdot \bmM_j \bc_j,
\eeq
where $\bmM_j$ is the dipole matrix for the corresponding defect, $\left.\nabla_{\bx} u^{(0)}\right|_{\bY_j}$ is the gradient of the displacement in the 
unperturbed configuration evaluated at the centre of the defect $\bY_j = (d_j\cos\varphi_j,d_j\sin\varphi_j)$, given by (\ref{nabla}), and 
\beq
\bc_j = \frac{1}{2d_j^{3/2}}\left[-\sin\frac{3\varphi_j}{2},\cos\frac{3\varphi_j}{2}\right].
\eeq

As a result, we obtain the formula
\beq
\label{phi}
\phi = -\frac{2}{A_\modIII^{(0)}} \sum_{j=1}^{3} \Delta K_\modIII^{(j)},
\eeq
which, together with (\ref{a30}) and (\ref{delk2}), allows for the analysis of the quasi-static propagation of the interfacial crack interacting with small 
micro-cracks and inclusions.

\section{Numerical results}
\label{sec4}

In this section, we show numerical results based on the explicit asymptotic formulae for the perturbation of stress intensity factor (\ref{delk2}) and for 
the crack advance (\ref{phi}). In Sec. \ref{sec41} we analyse the shielding and amplification effects on the crack-tip field of two defect arrangements 
(see Fig. \ref{fig02}). From this analysis it is possible to predict whether the crack will propagate (amplification) or not (shielding) from the specified 
position. In the case of propagation, the formula (\ref{phi}) allows us to incrementally calculate the crack advance, so that we can predict the extent of crack 
propagation and, in particular, whether the crack will arrest at a certain point or continue propagating. This is discussed in Sec. \ref{sec42}.

\subsection{Shielding and amplification effects.}
\label{sec41}

A neutral defect configuration is a configuration for which the perturbation of the stress intensity factor is zero, $\Delta K_\modIII = 0$. For non-neutral 
configurations, we may have {\it shielding} effect when the defects produce a decrease of the stress intensity factor, $\Delta K_\modIII/K_\modIII^{(0)} < 0$, 
or {\it amplification} effect in the opposite case, $\Delta K_\modIII/K_\modIII^{(0)} > 0$.

Before the discussion of computations, we derive simplified asymptotic formulae, which are valid in the case where the loading is applied far away from 
the crack tip. Let us denote the support of the applied loading by $\sup p_\pm \subset (-\infty,-a)$, $a > 0$. We have the following reduced asymptotic formula 
as $a \to +\infty$, valid for any elliptic inclusion with semi-axes $l_a, l_b$ and making an angle $\alpha$ with the positive $x_1$-direction
\beq
\barr{r}
\ds \frac{\Delta K_\modIII}{K_\modIII^{(0)}} \sim \frac{1}{2} \frac{l_a l_b}{d^2}(1 + e)(\mu_\star - 1)\frac{\mu_\mp}{\mu_+ + \mu_-}
\left\{\frac{1}{e + \mu_\star}\sin\left(\frac{3\varphi}{2} - \alpha\right)\sin\left(\frac{\varphi}{2} - \alpha\right) \right. \\[5mm]
\ds \left. + \frac{1}{1 + e\mu_\star}\cos\left(\frac{3\varphi}{2} - \alpha\right)\cos\left(\frac{\varphi}{2} - \alpha\right)\right\}.
\earr
\eeq

From this formula we can derive the limiting cases as $a \to +\infty$ for different type of defect, namely:

-- for a microcrack:
\beq
\label{mc}
\frac{\Delta K_\modIII}{K_\modIII^{(0)}} \sim \frac{1}{2} \frac{l^2}{d^2} \frac{\mu_\mp}{\mu_+ + \mu_-} 
\cos\left(\frac{3\varphi}{2} - \alpha\right)\cos\left(\frac{\varphi}{2} - \alpha\right);
\eeq

-- for a rigid line inclusion:
\beq
\label{rl}
\frac{\Delta K_\modIII}{K_\modIII^{(0)}} \sim -\frac{1}{2} \frac{l^2}{d^2} \frac{\mu_\mp}{\mu_+ + \mu_-} 
\sin\left(\frac{3\varphi}{2} - \alpha\right)\sin\left(\frac{\varphi}{2} - \alpha\right);
\eeq

-- for an elliptic void:
\beq
\frac{\Delta K_\modIII}{K_\modIII^{(0)}} \sim \frac{1}{2} \frac{l_a l_b}{d^2} (1/e + 1) \frac{\mu_\mp}{\mu_+ + \mu_-} 
\left\{e\sin\left(\frac{3\varphi}{2} - \alpha\right)\sin\left(\frac{\varphi}{2} - \alpha\right) + 
\cos\left(\frac{3\varphi}{2} - \alpha\right)\cos\left(\frac{\varphi}{2} - \alpha\right)\right\};
\eeq

-- for an elliptic rigid inclusion:
\beq
\frac{\Delta K_\modIII}{K_\modIII^{(0)}} \sim -\frac{1}{2} \frac{l_a l_b}{d^2} (1/e + 1) \frac{\mu_\mp}{\mu_+ + \mu_-} 
\left\{\sin\left(\frac{3\varphi}{2} - \alpha\right)\sin\left(\frac{\varphi}{2} - \alpha\right) + 
e\cos\left(\frac{3\varphi}{2} - \alpha\right)\cos\left(\frac{\varphi}{2} - \alpha\right)\right\};
\eeq

-- for a line defect with soft bonding and transmission conditions given by (\ref{trsoft}):
\beq
\frac{\Delta K_\modIII}{K_\modIII^{(0)}} \sim \frac{1}{2} \frac{l^2}{d^2} \frac{\mu_\mp}{\mu_+ + \mu_-} \frac{\kappa}{l + \kappa} 
\cos\left(\frac{3\varphi}{2} - \alpha\right)\cos\left(\frac{\varphi}{2} - \alpha\right);
\eeq

-- for a stiff line defect with transmission conditions given by (\ref{trstiff}):
\beq
\frac{\Delta K_\modIII}{K_\modIII^{(0)}} \sim -\frac{1}{2} \frac{l^2}{d^2} \frac{\mu_\mp}{\mu_+ + \mu_-} \frac{1}{1 + \kappa l} 
\sin\left(\frac{3\varphi}{2} - \alpha\right)\sin\left(\frac{\varphi}{2} - \alpha\right).
\eeq
Note that the formula (\ref{mc}) in the case of a homogeneous body was previously obtained by Gong (1995).

\begin{figure}[!htcb]
\begin{center}
\includegraphics[width=12cm]{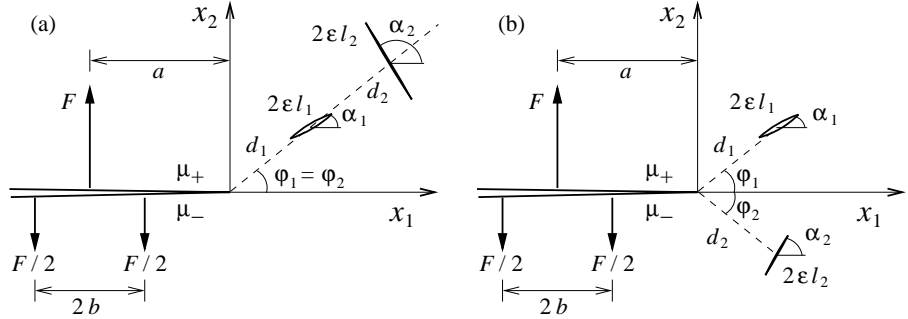}
\caption{\footnotesize Two neutral arrangements for remotely applied loading, $a \to +\infty$. 
Arrangement (a): a microcrack of length $2\varepsilon l_1$, orientation $\alpha_1$ and placed at the point 
$(d_1\cos\varphi_1,d_1\sin\varphi_1)$ and a rigid line inclusion of length $2\varepsilon l_2 = 4\varepsilon l_1$, orientation  
$\alpha_2 = \alpha_1 - \pi/2$ and placed at the point $(2d_1\cos\varphi_1,2d_1\sin\varphi_1)$.
Arrangement (b): a microcrack of length $2\varepsilon l_1$, orientation $\alpha_1$ and placed at the point 
$(d_1\cos\varphi_1,d_1\sin\varphi_1)$ and a rigid line inclusion of length $2\varepsilon l_2 = 2\varepsilon l_1 \sqrt{\mu_-/\mu_+}$, orientation 
$\alpha_2 = \pi/2 - \alpha_1$ and placed at the point $(d_1\cos\varphi_1,-d_1\sin\varphi_1)$.}
\label{fig02}
\end{center}
\end{figure}

From these simplified asymptotic formulae, it is possible to design neutral configurations, for the case where the loading is applied at large distance from 
the crack tip.  To illustrate this, we show in Fig. \ref{fig02} two examples of defects arrangements. The case of Fig.\ \ref{fig02}a consists of two defects, 
a microcrack, denoted by subscript 1, and a rigid line inclusion, denoted by subscript 2, located in the same half-plane and such that 
\beq
\label{confa}
l_1/d_1 = l_2/d_2, \quad \phi_1 - \phi_2 = 0, \quad \alpha_1 - \alpha_2 =  \pi/2.
\eeq
The case of Fig.\ \ref{fig02}b consists of the same two defects as above, a microcrack and a rigid line inclusion, but they are located in different half-planes 
and such that 
\beq
\label{confb}
\mu_- l_1^2 / d_1^2 = \mu_+ l_2^2 / d_2^2, \quad \phi_1 + \phi_2 = 0, \quad \alpha_1 + \alpha_2 = \pi/2.
\eeq
%%%%%%%%%%%%%%%%%%%%%%%%%%%%%%%%%%%%%%%%%%%%%%%%%%%%%%%%%%%%%%%%%%%%%%
%%%%%%%%%%%%%%%%%%%%%%%%%%%%%%%%%%%%%%%%%%%%%%%%%%%%%%%%%%%%%%%%%%%%%%

When the loading is applied at a finite distance from the crack tip, arrangements (a) and (b) shown in Fig.\ \ref{fig02} are, in general, not neutral any longer, 
and we may have shielding and amplification effects. To illustrate this, we consider a three-point load as shown in Fig. \ref{fig02}. This load is self-balanced 
for any $b < a$, and is symmetrical only for $b = 0$. 

\begin{figure}[!htcb]
\begin{center}
\includegraphics[width=10cm]{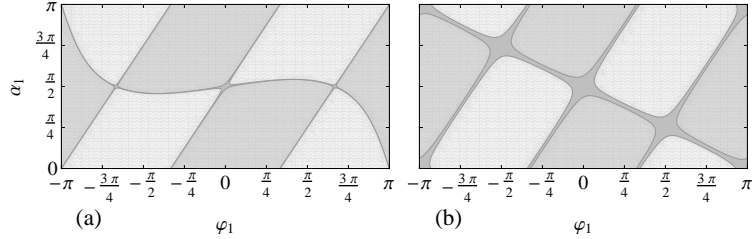}
\caption{\footnotesize Shielding (light grey), amplification (medium grey) and neutral (dark grey) regions created by a microcrack and a rigid line inclusion 
arranged as shown in Fig.\ \ref{fig02}a. The diagrams are for a weak interface between two identical materials, $\eta = 0$, and symmetrical loading, $b = 0$. 
The position of the microcrack is characterised by the angles $\varphi_1$ and $\alpha_1$, whereas the position of the rigid line inclusion is determined by 
conditions (\ref{confa}). The dark grey region corresponds to neutral configurations, that is configurations for which the perturbation is negligibly small 
(within the accuracy $\delta = 10^{-6}$). Figure (a): load is applied close to the crack tip, $a = 3$. Figure (b): load is applied at a large distance from the 
crack tip, $a = 100$.}
\label{figneut1eta0a}
\end{center}
\end{figure}

We consider first arrangement (a) for the case of a weak interface between two identical materials ($\mu_+ = \mu_- = \mu$) and symmetrical load, $b = 0$, see Fig.\ 
\ref{figneut1eta0a}. In the computation we use the values: $\varepsilon l_1 = 0.1$, $d_1 = 1$, $\varepsilon l_2 = 0.2$, $d_2 = 2$. On the 
diagrams, the horizontal axis stands for the angle $\varphi_1$ ($-\pi < \varphi_1 < \pi$) defining the angular position of the centre of the microcrack 1 with 
respect to the crack tip. On the vertical axis we report the value of the angle $\alpha_1$ ($0 < \alpha_1 < \pi$) defining the crack orientation with respect to the 
$x_1$-direction. The position of the rigid line inclusion 2 is determined by (\ref{confa}). It is worth noting that, in a real experiment, the perturbation of the 
stress intensity factor $\Delta K_\modIII$ can be measured with the accuracy $\delta = \Delta K_\modIII/K_\modIII^{(0)}$ with respect to the unperturbed value 
$K_\modIII^{(0)}$, so that the region corresponding to shielding is defined by $\Delta K_\modIII < -\delta K_\modIII^{(0)}$ and shadowed by light grey in the 
diagrams. The region corresponding to amplification is defined by $\Delta K_\modIII > \delta K_\modIII^{(0)}$ and shadowed by medium grey. The dark grey region 
in the diagrams corresponds to situations for which $|\Delta K_\modIII| < \delta K_\modIII^{(0)}$, that is the perturbation is negligibly small and the 
corresponding configuration could be regarded as neutral.

Fig. \ref{figneut1eta0a}a corresponds to a two-point (symmetrical) load situated close to the crack tip, $a = 3$. The diagram is symmetrical with 
respect to the angle $\varphi_1$, as the load is symmetrical and the material is homogeneous. As the distance $a$ of the load from the crack tip increases, 
the diagram changes and the dark grey region, corresponding to neutral configurations, enlarges, as shown in Fig. \ref{figneut1eta0a}b where we set $a = 100$.
%%%%%%%%%%%%%%%%%%%%%%%%%%%%%%%%%%%%%%%%%%%%%%%%%%%%%%%%%%%%%%%%%%%%%%
%%%%%%%%%%%%%%%%%%%%%%%%%%%%%%%%%%%%%%%%%%%%%%%%%%%%%%%%%%%%%%%%%%%%%%

\begin{figure}[!htcb]
\begin{center}
\includegraphics[width=12cm]{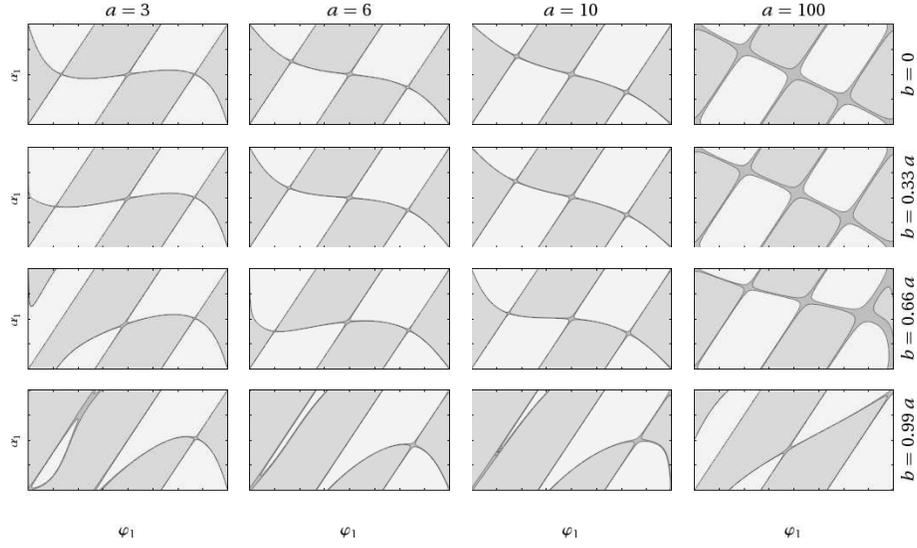}
\caption{\footnotesize Shielding (light grey), amplification (medium grey) and neutral (dark grey) regions created by a microcrack and a rigid line inclusion 
arranged as shown in Fig.\ \ref{fig02}a. The diagrams are for a weak interface between two identical materials, $\eta = 0$, and three-point loading, characterised 
by parameters $a$ and $b$. From left to right the distance $a$ of loading from the crack tip increases from $3$ to $100$. From up to bottom the asymmetry 
$b$ of loading increases from $0$ to $0.99 a$.}
\label{figneut1eta0b}
\end{center}
\end{figure}

In Fig.\ \ref{figneut1eta0b}, we analyse the effect of the asymmetry of loading for the same configuration as in Fig.\ \ref{figneut1eta0a}. The diagrams 
correspond to a three-point loading, characterised by the parameters $a$ and $b$, as shown in Fig. \ref{fig02}. The definition and range of the angles $\alpha_1$ 
and $\varphi_1$ are the same as in Fig.\ \ref{figneut1eta0a} ($-\pi < \varphi_1 < \pi$, $0 < \alpha_1 < \pi$). We consider four values of the parameter $a$, 
namely $a = 3,6,10,100$, from left to right. This allows us to trace different situations corresponding to the increasing distance of the loading from the crack tip
(keeping $b$ fixed). The asymmetry of loading is measured by the parameter $b$, which increases from the upper part (symmetrical loading $b = 0$) to the lower 
part ($b = 0.99 a$) of the figure. 
%%%%%%%%%%%%%%%%%%%%%%%%%%%%%%%%%%%%%%%%%%%%%%%%%%%%%%%%%%%%%%%%%%%%%%
%%%%%%%%%%%%%%%%%%%%%%%%%%%%%%%%%%%%%%%%%%%%%%%%%%%%%%%%%%%%%%%%%%%%%%

From the results presented in Fig.\ \ref{figneut1eta0b}, the following conclusions can be drawn. First, the neutral region (shadowed dark grey) enlarges when 
moving from the diagrams on the left to those on the right. This finding, already evident in Fig.\ \ref{figneut1eta0a} where only the values $a = 3$ and 
$a = 100$ were considered with $b = 0$, confirms the validity of the simplified formulae (\ref{mc}) and (\ref{rl}), established in the literature (Gong, 1995). 
Secondly, it is clear that the asymmetry of loading plays an important role. Indeed, even for $a = 100$ and $b = 0.33a$, or $b = 0.66a$, when the nearest point 
force is still well separated from the crack tip, the shielding--amplification diagrams are essentially different from those of the symmetrical case ($b = 0$). 
Thirdly, since the distance of the loading from the crack tip is measured by $a - b$, the neutral region shrinks by increasing the asymmetry $b$ at fixed $a$ 
(thus moving from the diagrams on the upper part to those on the lower part).

\begin{figure}[!htcb]
\begin{center}
\includegraphics[width=12cm]{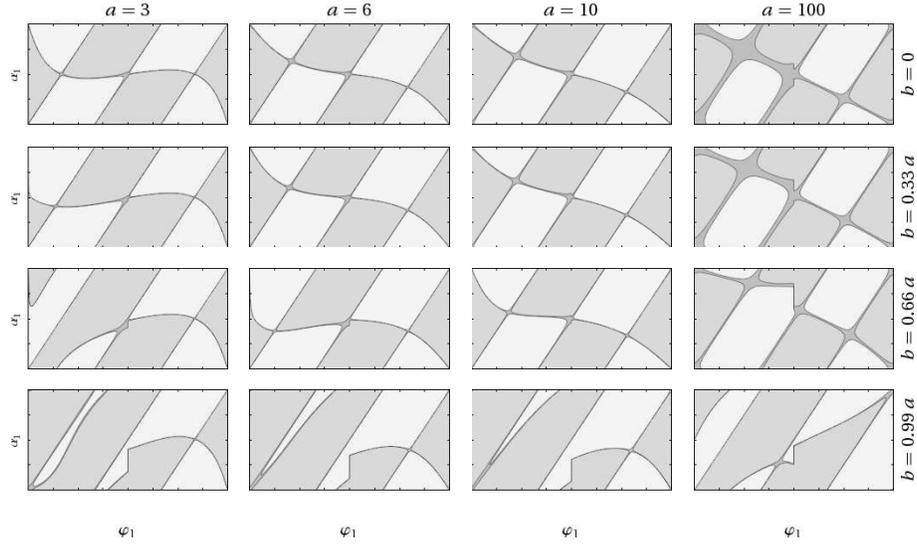}
\caption{\footnotesize Shielding (light grey), amplification (medium grey) and neutral (dark grey) regions created by a microcrack and a rigid line inclusion 
arranged as shown in Fig.\ \ref{fig02}a. The diagrams are for a bimaterial solid with $\eta = 0.67$, and three-point loading, characterised 
by parameters $a$ and $b$. From left to right the distance $a$ of loading from the crack tip increases from $3$ to $100$. From up to bottom the asymmetry 
$b$ of loading increases from $0$ to $0.99 a$.}
\label{figneut1eta0.67b}
\end{center}
\end{figure}

\begin{figure}[!htcb]
\begin{center}
\includegraphics[width=12cm]{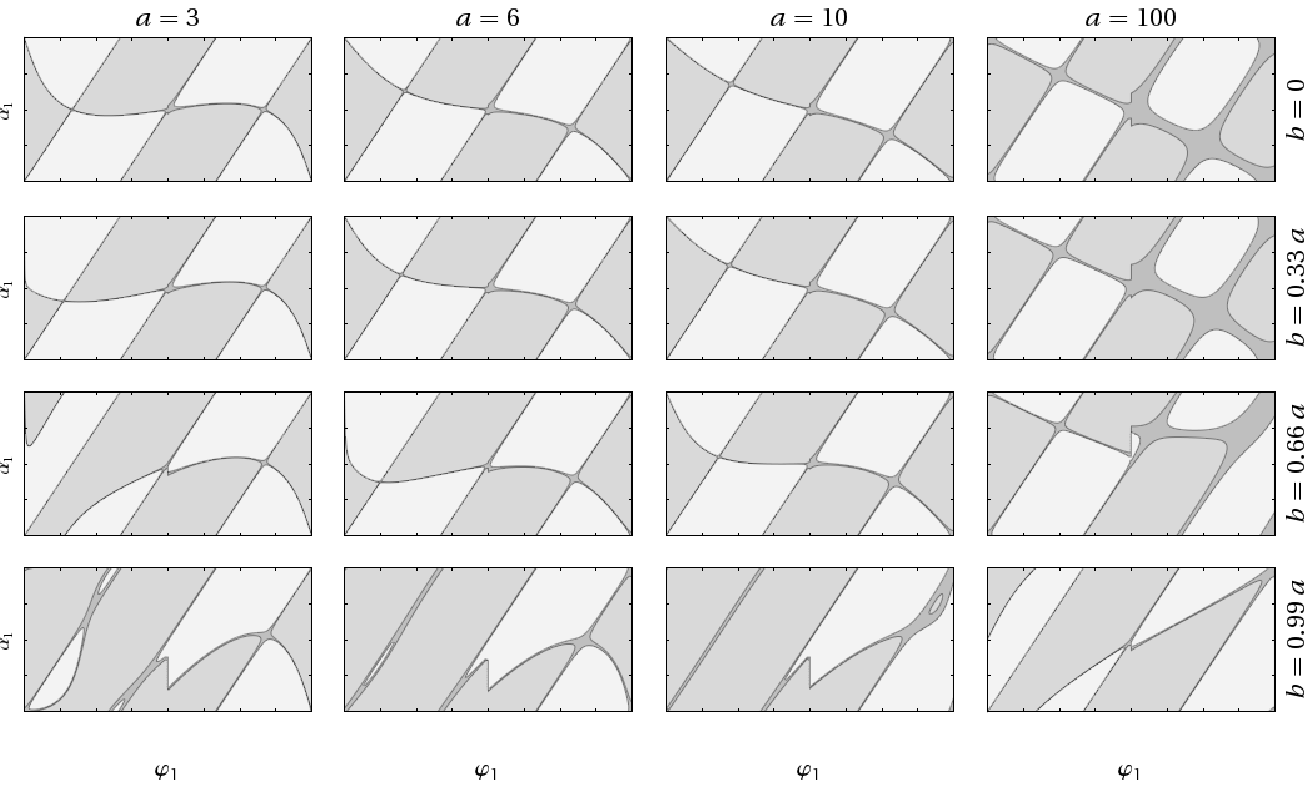}
\caption{\footnotesize Shielding (light grey), amplification (medium grey) and neutral (dark grey) regions created by a microcrack and a rigid line inclusion 
arranged as shown in Fig.\ \ref{fig02}a. The diagrams are for a bimaterial solid with $\eta = -0.67$, and three-point loading, characterised 
by parameters $a$ and $b$. From left to right the distance $a$ of loading from the crack tip increases from $3$ to $100$. From up to bottom the asymmetry 
$b$ of loading increases from $0$ to $0.99 a$.}
\label{figneut1eta-0.67b}
\end{center}
\end{figure}

In Fig.\ \ref{figneut1eta0.67b} and Fig.\ \ref{figneut1eta-0.67b}, we show the influence of the inhomogeneity on the shielding and amplification effects. 
For this purpose, we consider the aforementioned three-point loading for the case of a bimaterial plane with contrast parameter $\eta = 0.67$ (diagrams of 
Fig.\ \ref{figneut1eta0.67b}) and $\eta = -0.67$ (diagrams of Fig.\ \ref{figneut1eta-0.67b}). All other parameters ($a, b, \alpha_1$ and $\varphi_1$) are defined 
in the same manner as above.
%%%%%%%%%%%%%%%%%%%%%%%%%%%%%%%%%%%%%%%%%%%%%%%%%%%%%%%%%%%%%%%%%%%%%%
%%%%%%%%%%%%%%%%%%%%%%%%%%%%%%%%%%%%%%%%%%%%%%%%%%%%%%%%%%%%%%%%%%%%%%
%%%%%%%%%%%%%%%%%%%%%%%%%%%%%%%%%%%%%%%%%%%%%%%%%%%%%%%%%%%%%%%%%%%%%
%%%%%%%%%%%%%%%%%%%%%%%%%%%%%%%%%%%%%%%%%%%%%%%%%%%%%%%%%%%%%%%%%%%%%%

The influence of the inhomogeneity is clearly observable only for a highly pronounced non-symmetrical loading which is applied close to the crack tip. Furthermore, 
the shielding and amplification effects are more pronounced when the defects are located in the softer material (upper half-plane, $0 < \varphi_1 < \pi$, in 
Fig.\ \ref{figneut1eta0.67b}; lower half-plane, $-\pi < \varphi_1 < 0$, in Fig.\ \ref{figneut1eta-0.67b}).

\begin{figure}[!htcb]
\begin{center}
\includegraphics[width=12cm]{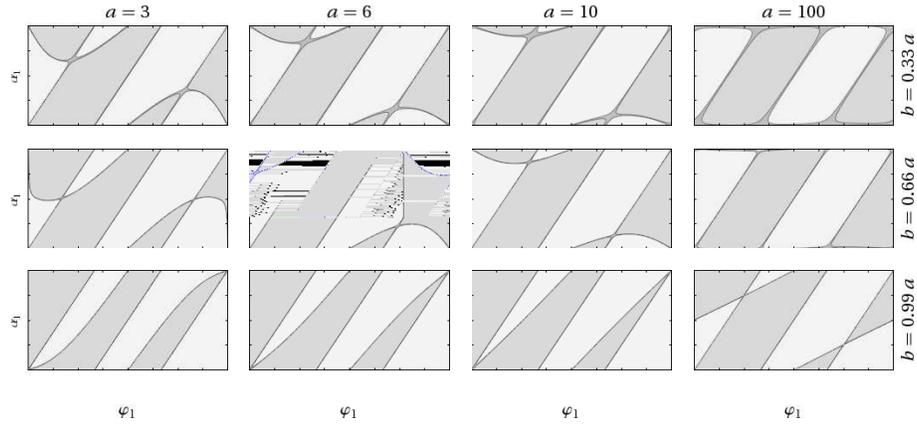}
\caption{\footnotesize Shielding (light grey), amplification (medium grey) and neutral (dark grey) regions created by a microcrack and a rigid line inclusion 
arranged as shown in Fig.\ \ref{fig02}b. The diagrams are for a weak interface between two identical materials, $\eta = 0$, and three-point loading, characterised 
by parameters $a$ and $b$. From left to right the distance $a$ of loading from the crack tip increases from $3$ to $100$. From up to bottom the asymmetry 
$b$ of loading increases from $0.33a$ to $0.99 a$..}
\label{figneut2eta0b}
\end{center}
\end{figure}

Finally, we comment on the arrangement (b), see Fig.\ \ref{fig02}b. This configuration turns out to be neutral for any symmetrical loading, 
independently of the distance $a$ from the crack tip, and the diagrams are very similar regardless of the bimaterial parameter $\eta$. For this reason, we 
present in Fig.\ \ref{figneut2eta0b} results only for the case of asymmetrical loading, $b \neq 0$, and weak interface between two identical materials, $\eta =0$.
%%%%%%%%%%%%%%%%%%%%%%%%%%%%%%%%%%%%%%%%%%%%%%%%%%%%%%%%%%%%%%%%%%%%%%
%%%%%%%%%%%%%%%%%%%%%%%%%%%%%%%%%%%%%%%%%%%%%%%%%%%%%%%%%%%%%%%%%%%%%%

It follows from Fig.\ \ref{figneut2eta0b} that the neutrality property of arrangement (b) is broken by the asymmetry of loading and thus amplification--shielding 
regions appear for $b \neq 0$. Moreover, the influence of asymmetry is not only quantitative but also qualitative, since the diagrams are completely different 
for different values of $b$. This emphasises the importance of the skew-symmetric weight functions (Piccolroaz et al., 2009) which are used to evaluate the 
contribution of the skew-symmetric load.

\clearpage

\subsection{Crack propagation and arrest}
\label{sec42}

In this section we discuss the influence of small defects on the crack propagation, in particular on the crack ``acceleration'' and arrest. Given a 
configuration of defects and position of the crack tip with respect to the defects, the formula (\ref{phi}) allows for the computation of the incremental crack 
advance $\phi$. It is possible then to update the configuration with the new position of the crack tip with respect to the defects and recompute the incremental 
crack advance in the new configuration, following an iterative procedure. The crack ``accelerates'' when the increment $\phi$ is increasing and ``decelerates'' 
in the opposite case. The total crack elongation is computed as 
\beq
x(N) = \sum_{i=0}^{N} \phi_i,
\eeq
where $N$ is the number of iterations.

\begin{figure}[!htcb]
\begin{center}
\includegraphics[width=10cm]{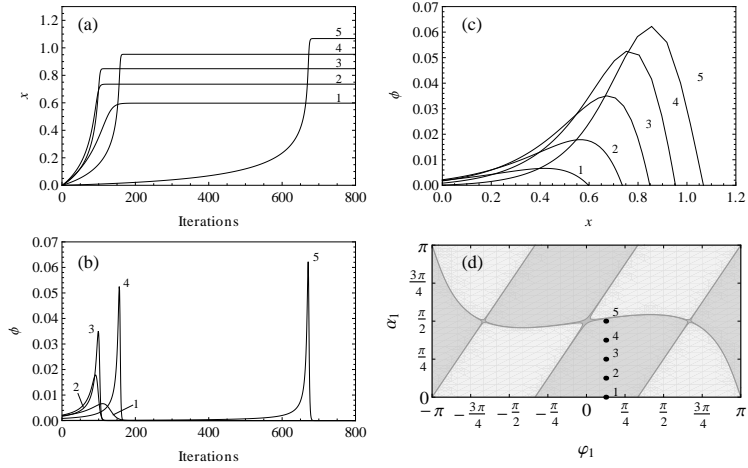}
\caption{\footnotesize Acceleration and arrest of the main crack along a weak interface between two identical materials, produced by a microcrack and a rigid line 
inclusion arranged as in Fig.\ \ref{fig02}a. The initial position of the microcrack relative to the crack tip is characterised by the values 
$d_1 = 1$, $\varphi_1 = \pi/8$. Five cases of angular orientation of the microcrack are considered, namely $\alpha_1 = 0,\pi/8,\pi/4,3\pi/8,\pi/2$, 
and denoted by labels from 1 to 5. The initial position and the inclination of the rigid line inclusion is determined by conditions (\ref{confa}).
The crack elongation, denoted by $x$, is assumed equal to 0 in the initial configuration. Figure (a): crack elongation $x$ vs.\ number of iterations. 
Figure (b): increment $\phi$ vs.\ number of iterations. Figure (c): increment $\phi$ vs.\ crack elongation $x$. Figure (d): initial configurations are 
indicated by spots in the shielding--amplification diagram.}
\label{figprop1stop}
\end{center}
\end{figure}

In Fig.\ \ref{figprop1stop} we show the numerical results for the case of two defects (a microcrack and a rigid line inclusion) arranged as shown in Fig.\ 
\ref{fig02}a. The body is assumed homogeneous $\eta = 0$ and the loading is symmetric $b = 0$. The initial position of the microcrack 
relative to the main crack tip is characterised by the values $d_1 = 1$, $\varphi_1 = \pi/8$. Five cases of angular orientation of the microcrack are 
considered, namely $\alpha_1 = 0,\pi/8,\pi/4,3\pi/8,\pi/2$, and denoted by labels from 1 to 5, see Fig.\ \ref{figprop1stop}d. The initial position and the 
inclination of the rigid line inclusion is determined by conditions (\ref{confa}). 

Fig. \ref{figprop1stop}a shows the crack elongation $x$ as a function of the number of iterations. For all cases, we may observe 
crack acceleration followed by a rapid deceleration and arrest, see Fig.\ \ref{figprop1stop}b. The crack arrests when a neutral configuration is reached at that 
particular position of the crack tip.. 
The total elongation at arrest is different for different cases. In the case denoted by 5, the crack advance is initially very slow, because the initial 
configuration is close to the border of the neutral region (dark grey region in Fig.\ \ref{figprop1stop}d), but the crack reaches the largest elongation at arrest. 
The incremental crack advance $\phi$ is plotted against the crack elongation $x$ in Fig.\ \ref{figprop1stop}c.
%%%%%%%%%%%%%%%%%%%%%%%%%%%%%%%%%%%%%%%%%%%%%%%%%%%%%%%%%%%%%%%%%%%%%%
%%%%%%%%%%%%%%%%%%%%%%%%%%%%%%%%%%%%%%%%%%%%%%%%%%%%%%%%%%%%%%%%%%%%%%

\begin{figure}[!htcb]
\begin{center}
\includegraphics[width=10cm]{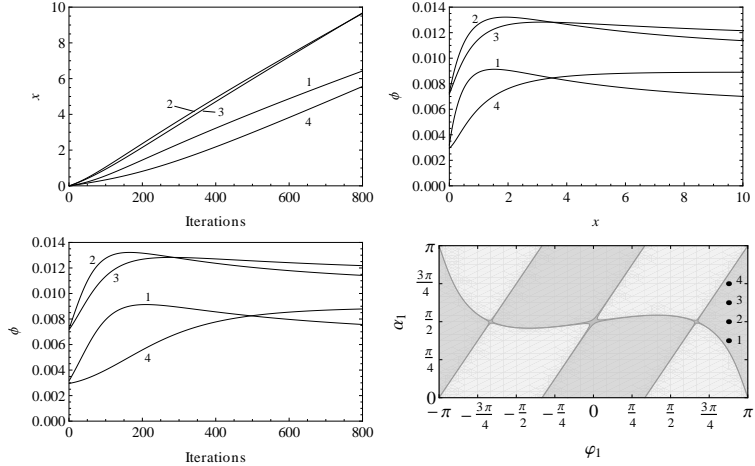}
\caption{\footnotesize Initial acceleration followed by steady-state propagation of the main crack along a weak interface between two identical materials, produced 
by a microcrack and a rigid line inclusion arranged as in Fig.\ \ref{fig02}a. The initial position of the microcrack relative to the crack tip 
is characterised by the values $d_1 = 1$, $\varphi_1 = 7\pi/8$. Four cases of angular orientation of the microcrack are considered, namely 
$\alpha_1 = 3\pi/8,\pi/2,5\pi/8,3\pi/4$, and denoted by labels from 1 to 4. The initial position and the inclination of the rigid line inclusion is determined 
by conditions (\ref{confa}). The crack elongation, denoted by $x$, is assumed equal to 0 in the initial configuration. Figure (a): crack elongation $x$ vs.\ 
number of iterations. Figure (b): increment $\phi$ vs.\ number of iterations. Figure (c): increment $\phi$ vs.\ crack elongation $x$. Figure (d): initial 
configurations are indicated by spots in the shielding--amplification diagram.}
\label{figprop1acc}
\end{center}
\end{figure}

In Fig.\ \ref{figprop1acc} we consider the same situation as in Fig.\ \ref{figprop1stop}, however now the initial position of the microcrack relative to the 
main crack tip is characterised by the values $d_1 = 1$, $\varphi_1 = 7\pi/8$, and four angular orientation are considered, 
$\alpha_1 = 3\pi/8,\pi/2,5\pi/8,3\pi/4$, see Fig.\ \ref{figprop1acc}d. We may observe a crack acceleration followed by a slow deceleration until a steady-state 
propagation is reached. In fact, as the crack is propagating and the angle $\varphi_1$ is increasing (up to the maximum value $\pi$), the corresponding 
configuration will always remain in the amplification region, see Fig.\ \ref{figprop1acc}d. 
%%%%%%%%%%%%%%%%%%%%%%%%%%%%%%%%%%%%%%%%%%%%%%%%%%%%%%%%%%%%%%%%%%%%%%
%%%%%%%%%%%%%%%%%%%%%%%%%%%%%%%%%%%%%%%%%%%%%%%%%%%%%%%%%%%%%%%%%%%%%%

\section{Conclusions}
\label{sec5}

In the present paper we show that the new weight functions for a bimaterial plane constructed in Piccolroaz et al.\ (2009) together with the 
dipole matrix approach provide an efficient tool in analysis of perturbation problems arising from the interaction of an interfacial crack with small 
defects, including elastic inclusions, microcracks, rigid line inclusions, and line defects with imperfect bonding.
We also show that the skew-symmetric weight functions play a crucial role as they permit to evaluate the contribution of the skew-symmetric 
load applied along the crack faces.

We illustrate the method with a number of applications, ranging from the analysis of shielding--amplification effects of the defects on the crack-tip 
fields to the perturbation modelling of the quasi-static interface crack propagation. The influence of asymmetry and position of the loading with respect to 
the crack tip is discussed and quantified. 

As a final remark, we note that the method can be extended to the case when the number of defects becomes increasingly high (clouds of defects). However, 
this requires additional accurate treatment of the interaction between defects and it is beyond the scope of the present paper.

\vspace{10mm}
{\bf Acknowledgement.} This research was supported by the Research-In-Groups (RiGs) programme of the International Centre for Mathematical 
Sciences, Edinburgh, Scotland. In addition, A.P. and G.M. gratefully acknowledge the support from the European Union Seventh Framework Programme 
under contract numbers PIEF-GA-2009-252857 and PIAP-GA-2009-251475, respectively.

\clearpage
%%%%%%%%%%%%%%%%%%%%%%%%%%%%%%%%%%%%%%%%%%%%%%%%%%%%%%%%%%%%%%%%%%%%%%%
%Appendix
\appendix
\renewcommand{\theequation}{\thesection.\arabic{equation}}

\section{APPENDIX}
\setcounter{equation}{0}

\subsection{The dipole matrix for an elastic elliptic inclusion in antiplane shear}
\label{app01}

Although the general problem of an elliptic inclusion in an infinite medium subject to remote uniform stress has been known since the mid of the last 
century (Donnell, 1941; Hardiman, 1954), the dipole matrix in the case of antiplane shear is practically impossible to find in the existing literature. For 
this reason, we have decided to present here not only the matrix but also in brief its derivation.

We consider an elastic elliptic inclusion $g$, with shear modulus $\mu_0$, embedded in an infinite body with shear modulus $\mu$, undergoing 
antiplane shear deformation. The boundary of the inclusion is denoted by $\partial g = \{(x,y): x^2/a^2 + y^2/b^2 = 1\}$.

The problem is formulated as follows:
$$
\Delta u = 0 \quad \text{in} \quad \Reals^2 \setminus \overline{g},
$$
$$
\Delta u_0 = 0 \quad \text{in} \quad g,
$$
$$
u = u_0, \quad \mu \frac{\partial u}{\partial n} = \mu_0 \frac{\partial u_0}{\partial n} \quad 
\text{on} \quad \partial g,
$$
\beq
\label{bvp}
u = B_R x - B_I y + O(1) \quad \text{as} \quad r = \sqrt{x^2 + y^2} \to \infty,
\eeq
where $u_0$ and $u$ are the displacements in the interior and exterior of the inclusion, respectively, $\bn$ is an outward unit normal on $\partial g$ and 
$B_R$, $B_I$ are known constants. 

The displacement in the inclusion is linear (Horgan, 1995; Ru and Schiavone, 1996), so that
\beq
u_0 = A_R x - A_I y \quad \text{in} \quad g.
\eeq

Our objective is to compute the next term in the asymptotic of the displacement field at infinity, eq. (\ref{bvp}), which will be provide us with the required 
dipole matrix. To this purpose we will use the theory of analytic functions. Thus, we write $u_0(x,y) = \Re U^0(z) = U^0_R(z)$ and 
$u(x,y) = \Re U(z) = U_R(z)$, where $U^0 = Az$ is a linear function in $g$ and $U$ is an analytic function in $\Omega \setminus \overline{g}$ satisfying the 
following conditions
$$
U = Bz + \frac{D_1}{z} + O(|z|^{-2}), \quad B = B_R + i B_I, \quad D = D_R + i D_I, \quad z \to \infty,
$$
\beq
\label{compl}
U^0_R - U_R = 0, \quad \mu_0 U^0_I - \mu U_I = 0 \quad \text{on} \quad \partial g.
\eeq
Note that, in order to find the dipole matrix, we do not need the complete solution but only the constant $D_1$. Furthermore, we introduce the conformal 
mapping 
$$
z = c\left( \zeta + \frac{\lambda}{\zeta} \right), \quad c > 0, \quad 0 < \lambda < 1,
$$
which maps the ellipse $\partial g$ in the $z$-plane onto the unit circle in the $\zeta$-plane. The segment $d$ connecting the foci of the ellipse is mapped 
onto the circle $|\zeta| = \sqrt{\lambda}$, so that the region $g \setminus d$ is transformed into the ring $\sqrt{\lambda} < |\zeta| < 1$.

We seek for the solution in the transformed coordinates in the form 
\beq
\label{transformed}
U^0 = Ac(\zeta + \lambda/\zeta), \quad U = \beta \zeta + \gamma_1/\zeta.
\eeq
Such a solution exists with constants $\beta$ and $\gamma_1$ satisfying the following conditions
$$
\gamma_{1R} = \beta_R \frac{\mu(1 + \lambda) - \mu_0(1 - \lambda)}{\mu(1 + \lambda) + \mu_0(1 - \lambda)}, \quad 
\gamma_{1I} = -\beta_I \frac{\mu(1 - \lambda) - \mu_0(1 + \lambda)}{\mu(1 - \lambda) + \mu_0(1 + \lambda)}. 
$$
Going back to the original coordinates, from the asymptotic expansion $\zeta = z/c - c\lambda/z + O(|z|^{-3})$, $z \to \infty$, we get
$$
\beta = cB, \quad D_1 = \gamma_1 c - \beta \lambda c.
$$
As a result, we obtain the required two terms asymptotic of the displacement at infinity in the form
\beq
u = B_R x - B_I y - \frac{1}{2\pi} \{B_R,-B_I\} \cdot \bP\, \frac{\{x,y\}}{x^2 + y^2} + 
O(r^{-2}), \quad \text{as} \quad r \to \infty,
\eeq
where, taking into account that $c = (a + b)/2$ and $\lambda = (a - b)/(a + b)$, the dipole matrix $\bP$ is given by
\beq
\bP = -\pi ab(a + b)
\left[
\barr{ll}
\ds \frac{\mu - \mu_0}{\mu a + \mu_0 b} & 0 \\[3mm]
0 & \ds \frac{\mu - \mu_0}{\mu_0 a + \mu b} 
\earr
\right].
\eeq

We obtain the solution we need in Sec. \ref{sec311} by defining new functions $\tilde{u} = u -B_R x + B_I y$ and $\tilde{u}_0 = u_0 -B_R x + B_I y$. 
Then $\tilde{u}$ and $\tilde{u}_0$ satisfy a BVP similar to (\ref{bvp}), in which the condition at infinity is replaced by $\tilde{u} \to 0$ as $r \to \infty$, 
and the ideal transmission conditions are replaced by imperfect transmission conditions as follows
$$
\tilde{u} = \tilde{u}_0, \quad 
\mu \frac{\partial \tilde{u}}{\partial n} - \mu_0 \frac{\partial \tilde{u}_0}{\partial n} = 
- (\mu - \mu_0) \bn \cdot \{B_R,-B_I\} \quad \text{on} \quad \partial g.
$$

\section{Asymptotics at infinity for a movable rigid line inclusion}
\label{app02}
One can find, as a particular case from Muskhelishvili (2008), a solution $U$ for analytic function in the complex plane with given conditions along the symmetric 
segment of the length $2a$ situated along the real $x$-axis: $U^+ = F(x)$, $U^- = F(x)$. The bounded solution for this problem reads 
\beq
U(z) = \frac{\sqrt{R(z)}}{\pi i} \int_{-a}^{a} \frac{F(t) dt}{\sqrt{R(t)}(t - z)},
\eeq
where $z = x + iy$, and $R(z) = z^2 - a^2$.
This solution vanishes at infinity if the following condition is satisfied
\beq
\int_{-a}^{a} \frac{F(t)}{\sqrt{-R(t)}} dt = 0.
\eeq
It is clear from (\ref{w3}) that this condition is satisfied, since for a rigid line inclusion $F(t) = Bt$, where 
$B = - \nabla_{\bx} u^{(0)}|_{\bY_3} \cdot \bPhi^T(\alpha_3) \be_1$, with $\bPhi$ being the rotation matrix and $\be_1$ the unit vector along the $x$-axis.
The asymptotics at infinity for this function takes the form
\beq
U(z) = \frac{Ba^2}{2z} + O(z^{-3}), \quad \text{as} \quad z \to \infty.
\eeq
Thus, the boundary layer solution for a rigid line inclusion in the coordinate system $\bxi_3$ is given by a formula similar to (\ref{boundary_layer_1}) with 
$\bmM_1$ replaced by (\ref{rigid-line}).

\end{document}